\documentclass[prl,aps,nofootinbib,showpacs,10pt,twocolumn]{revtex4}
\usepackage{graphicx,epsfig}

\begin{document}

\title{Hunting for a scalar glueball in exclusive $B$ decays}
\author{Cai-Dian L\"u$^{a}$, Ulf-G. Mei{\ss}ner$^{b,c}$, Wei Wang $^b$ and Qiang Zhao$^{a}$}
\affiliation{
 $^a$ Institute of High Energy Physics and Theoretical Physics Center for Science Facilities,
Chinese Academy of Sciences, Beijing 100049, China \\
 $^b$ Helmholtz-Institut f\"ur Strahlen- und Kernphysik and Bethe Center for Theoretical Physics,
Universit\"at Bonn, D-53115 Bonn, Germany\\
$^c$ Institute for Advanced Simulation, Institut f\"ur Kernphysik and J\"ulich Center for Hadron Physics,
JARA-FAME and JARA-HPC, Forschungszentrum J\"ulich, D-52425 J\"ulich, Germany}

\begin{abstract}
Using  flavor SU(3) symmetry for the light quarks validated by
the available experimental data, we propose an intuitive way to hunt
for a scalar glueball in $B$ decays. In the presence of mixing
between the glueball  and ordinary scalar mesons, we explore the
extraction of the mixing parameters. In particular, we discuss the
implication from the recently available experimental data and show  the sensitivities of
$B$ decays as a probe to the scalar structures. The future Super KEKB factory would
allow access to establishing the mixing pattern among  the
scalars, and possibly allow one to disentangle the long-standing puzzle
concerning the existence and mixings of the scalar
glueball predicted by QCD.
\end{abstract}

\maketitle


{\it Introduction.}
One has the most exotic forms of matter consistent with QCD are glueballs. These are bound states made of  the color
force carriers only.
Lattice QCD simulations  have suggested the mass of the
lowest-lying scalar glueball  around
1.5-1.8~GeV~\cite{Bali:1993fb,Chen:2005mg}. 
Despite several possible
candidates in this mass region,  the existence  of a scalar glueball
state is still under debate, largely because of the fact that the
lowest-lying scalar glueball has the same quantum numbers as the QCD vacuum,
and thereby mixes with  ordinary  quark-antiquark states.

Most glueball studies available in the literature have
focused on decay properties  and the production  in low-energy
processes. In fact, the study of the production  in $B$ decays is
another powerful way  to uncover the mysterious structure of scalar
mesons and figure out the gluon component inside~\cite{He:2006qk}.

The motivation of this work is to provide an up-to-date analysis of $B$
decays into a scalar meson plus a $J/\psi$,  particularly  in the
light of the recent data on $B/B_s\to J/\psi \pi^+\pi^-/K^+K^-$
decays from the LHCb, Belle and D0
collaborations~\cite{Li:2011pg,LHCb:2012ae,Aaij:2011ac,Abazov:2012dz}.
In view of these, we will suggest an intuitive way for the identification
of a glueball.

In the following, we shall consider  three scalar mesons
$f_0(1370)$,$f_0(1500)$ and $f_0(1710)$  all having  potentially
a large glue content, see  for instance
Refs.~\cite{Close:2000yk,Cheng:2006hu}. These three mesons, together with the isotriplet $a_0(1450)$ and isodoublet $K^*_0(1430)$ can form an SU(3) octet made of $\bar qq$, with one additional state arising from the mixing with the glueball~\cite{note_on_scalar_mesons_below_2GeV}.  From this viewpoint, without loss of
generality, the isosinglet scalar meson among $f_0(1370)$,$f_0(1500)$ and $f_0(1710)$ is expanded 
\begin{eqnarray}
 |f_0\rangle &=& \alpha_1| G\rangle +\alpha_2|\bar ss\rangle +\alpha_3 |\bar nn\rangle, \label{eq:mixing_coefficient}
\end{eqnarray}
in which the coefficient  $\alpha_1$ is the measure of the glue content.
The three coefficients satisfy  the unitarity condition $\alpha_1^2+\alpha_2^2+\alpha_3^2=1$.
Here, $n$ denotes the light quark flavors up and down and we work in the
isospin limit in what follows.

\begin{figure}
\begin{center}
\includegraphics[scale=0.6]{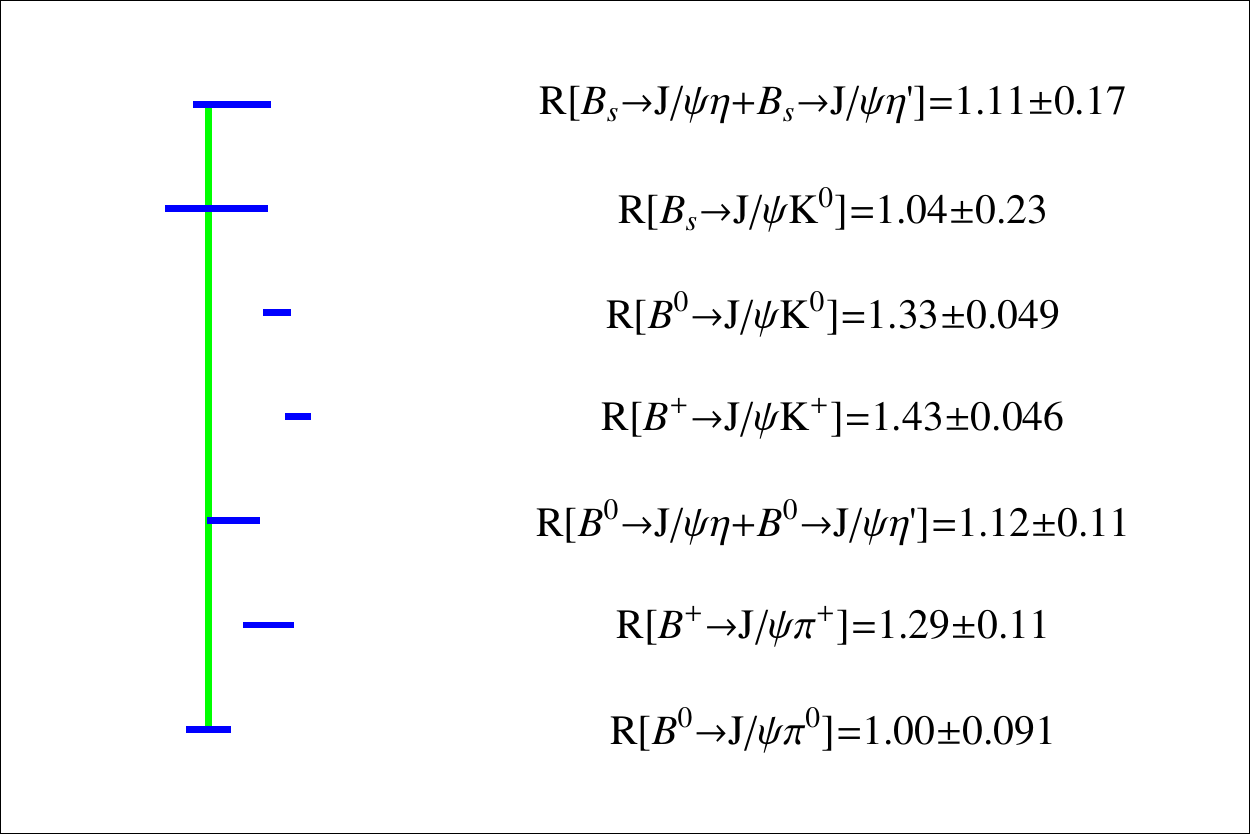}
\includegraphics[scale=0.6]{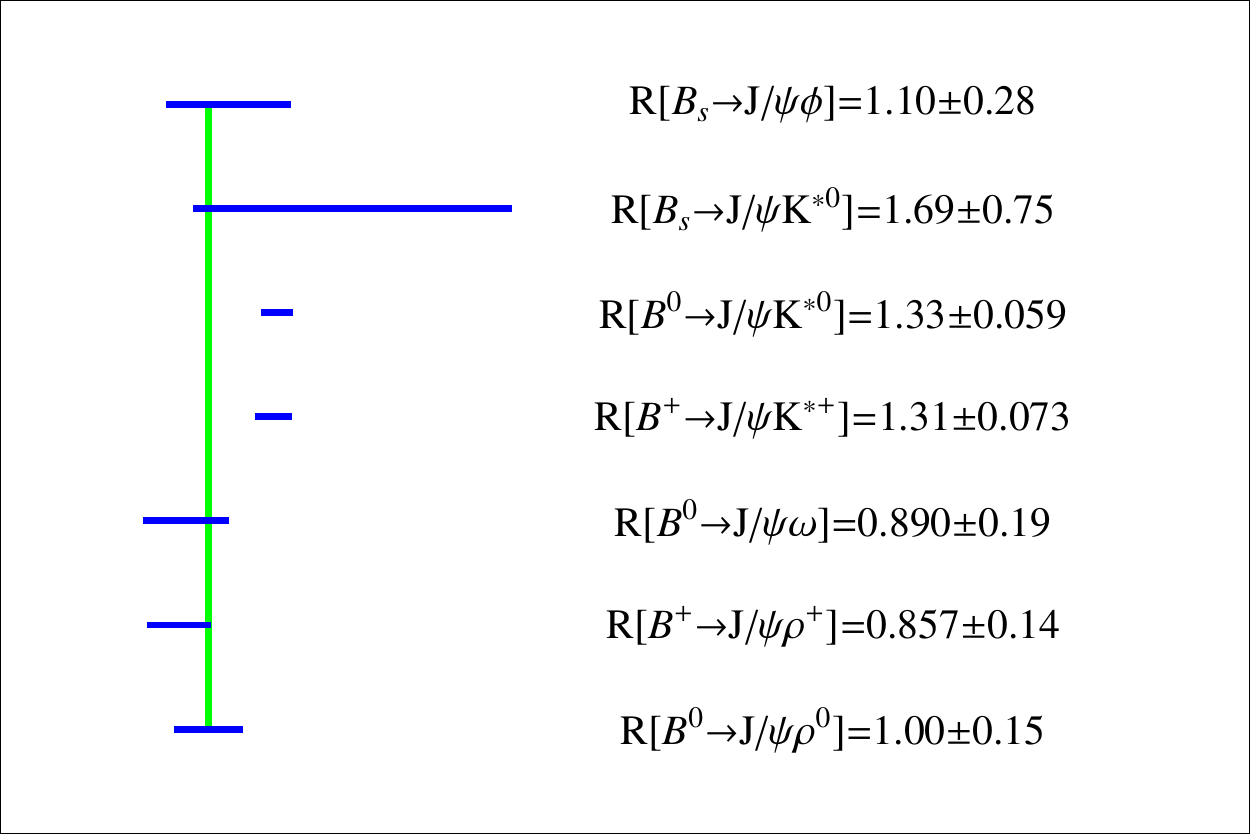}
\end{center}
\caption{Ratios of branching fractions of $B\to J/\psi P$ and $B\to J/\psi
  V$. The vertical lines denotes unity and correspond to the SU(3) symmetry limit.}
\label{fig:ratio}
\end{figure}

{\it General analysis based on SU(3) symmetry.}
To start, we will assume flavor SU(3) symmetry for the light $u,d,s$ quarks 
in the $B\to J/\psi M$ decay amplitudes. This symmetry has been partly tested 
in a few $B\to J/\psi P$ and $B\to J/\psi V$ processes~\cite{Colangelo:2010wg} 
and a good agreement with the data is found.  The underlying nature of these 
processes at the quark level is governed by the $b\to \bar cc s$ or $b\to \bar 
cc d$ transitions whose matrix elements can be related  using the flavor
symmetry.  For a better comparison to be made in the following we define the ratio:
\begin{eqnarray}
R[B_q\to J/\psi P_{q\bar q'}] &=& \frac{|V_{cd}|^2} {|V_{cq'}|^2}
\frac{|C_{\pi^0}|^2}{|C_{P}|^2}  
\frac{  \tau(B^0) } {  \tau(B_q)} \nonumber\\
&& \times \frac{ {\cal B}(B_q\to J/\psi P_{q\bar q'})}{{\cal B}(B^0\to J/\psi \pi^0) },
\end{eqnarray}
whose deviation from unity directly arises from the SU(3) symmetry
breaking effects. In this ratio, $C_{P}=1$ (except for  $-C_{\pi^0}=
C_{\eta_{q\bar q}}=1/\sqrt 2$) is the flavor wave function factor.
The ratios for the $B\to J/\psi V$ processes  are defined in a
similar way. Using the relevant experimental data for the branching
fractions of the various
channels~\cite{Beringer:1900zz,LHCb:2012cw,Chang:2012gnb}, we
collect the results for these ratios in Fig.~\ref{fig:ratio}.  The
vertical lines in this figure denotes unity  and thus corresponds
to the exact SU(3) symmetry limit.

Three observations can be made from the results presented in
Fig.~\ref{fig:ratio}. Firstly,  the current uncertainties in the $B_s$
decays are large but may get  significantly reduced  due to the
large amount of data accumulated from the LHC and future Super B
factories. Secondly, the SU(3) symmetry holds well in the $b\to d$
processes (namely $B\to (J/\psi \pi, J/\psi \eta^{(')})$ and 
$B_s\to J/\psi K$), and as well as in the $b\to s$ transition 
($B\to J/\psi K$ and $B_s\to J/\psi \eta^{(')}$).  Last, the excess
of the branching ratios for the $b\to s$ processes, roughly $30\%$, is
the same in both $B\to J/\psi P$ and $B\to J/\psi V$ decays.

After validating the flavor SU(3) symmetry, we now  explore the
consequences in the application to  $B\to J/\psi S$ decays.
Suppose in the near future we are equipped with the following
experimental data
\begin{eqnarray}
 {\cal B}_1&=& {\cal B}(B^0\to J/\psi K^*_0(1430)),\nonumber\\
 {\cal B}_2&=& {\cal B}(B^0_s\to J/\psi K^*_0(1430)),\nonumber\\
 {\cal B}_3&=& {\cal B}(B^0_s\to J/\psi f_0),\nonumber\\
 {\cal B}_4&=& {\cal B}(B^0\to J/\psi f_0),
\end{eqnarray}
where the second quantity can also be replaced by ${\cal B}(B^-\to J/\psi a_0^{-}(1450))$.
The first and third processes are induced by the $b\to s$
transition, while the other two arise from the $b\to d$ transition. In
the SU(3) symmetry  (to be more specific the U-spin symmetry which interchanges the $s$ and $d$ quarks) limit, ${\cal B}_1= {\cal B}_2$, but in order to
account for the symmetry breaking effects that can reach 30\% as we
have shown, we will retain the differences in $ {\cal B}_1$ and $
{\cal B}_2$. This treatment will refine our analysis based on the SU(3) symmetry and     greatly 
reduce the systematic errors in the analysis.


In the leading-Fock-state approximation, a scalar glueball is
composed of two constituent gluons. In exclusive $B$ decays, the two
gluons can be emitted from either the heavy $b$ quark or the light
quark. The emission of a collinear gluon from the heavy $b$ quark is
suppressed by $1/m_b^2$. Since the initial state does not contain
any valence gluon, in order to generate the glueball the quarks have
to be annihilated via the QCD interaction. Compared to the form
factor of $B$ to an ordinary $\bar qq$ transition, such a contribution
is suppressed by $\alpha_s(m_b\Lambda_{QCD})$, where the scale in
$\alpha_s$ has been set to the typical scale in the transition.
The  calculation 
in the perturbative QCD approach shows
that the $B$-to-glueball form factor is suppressed by a factor of
6-10~\cite{Wang:2009rc,Li:2008tk}.

In the following discussion, we will neglect the small contributions
from the glueball content, and thus only the $\bar nn$ ($\bar ss)$
component will contribute in $B$ ($B_s$) decays into a scalar meson plus
a $J/\psi$. As an important consequence, ${\cal B}_3/{\cal B}_1=
\alpha_2^2$ and $2{\cal B}_4/{\cal B}_2= \alpha_3^2$, while  the
product of ratios $1-{\cal B}_3/{\cal B}_1 -2{\cal B}_4/{\cal B}_2$
directly  reflects the size of the glueball component. {\it Any
significant deviation of $1-{\cal B}_3/{\cal B}_1 -2{\cal B}_4/{\cal
B}_2$ from 0 will be a  clear signal for a glueball.  }

 {\it   Implications from the present data.}
In Ref.~\cite{Li:2011pg}, the Belle Collaboration reported the
observation of a scalar mesonic state  $f_X$ from the process
$B_s\to J/\psi f_X\to  J/\psi \pi^+\pi^-$ with a significance  of
$4.2\sigma$:
\begin{eqnarray}
{\cal B} (B_s\to J/\psi f_X\to J/\psi \pi^+\pi^-)=(0.34^{+0.14}_{-0.15})\times 10^{-4}.
\end{eqnarray}
The mass and width of this resonance are determined as
\begin{eqnarray}
 m_{f_X}=(1.405\pm0.015^{+0.001}_{-0.007}) {\rm GeV},\nonumber\\
 \Gamma_{f_X}= (0.054\pm 0.033^{+0.014}_{-0.003}){\rm GeV} .
\end{eqnarray}
Subsequently, the LHCb collaboration has found a similar resonance 
\begin{eqnarray}
 m_{f_X}=(1.4751\pm0.0063) {\rm GeV},\nonumber\\
 \Gamma_{f_X}= (0.113\pm 0.011){\rm GeV}.
 \end{eqnarray}
The branching ratio (BR) of $B_s\to J/\psi f_X$ is  roughly $4\%$ of the
BR for $B_s\to J/\psi \phi$. Remembering that  ${\cal B}(B_s\to
J/\psi\phi) = (1.09^{+0.28}_{-0.23})\times 10^{-3}$~\cite{Beringer:1900zz}, we note that the
LHCb result   is consistent  with the Belle measurement.


The measured  branching ratio of $B_s\to J/\psi f_X \to J/\psi
\pi^+\pi^-$  is helpful to determine the mixing coefficient
$\alpha_2$ in $f_X$ together with
\begin{eqnarray}
 {\cal B}(B^0\to J/\psi K^0) &=& (8.71\pm 0.32)\times 10^{-4}.
\end{eqnarray}
Under the assumption of  factorization, we extract the coefficient $\alpha_2$ as
\begin{eqnarray}
 \alpha_2 &=& (0.27\pm 0.13) \times \frac{F_1^{B\to K}(m_{J/\psi}^2)}{0.53}   \nonumber\\
 &&\times \frac{1.22}{F_1^{B_s\to f_0(\bar ss)}(m_{J/\psi}^2)} \times \sqrt{\frac{10\%}{{\cal B}({f_X\to \pi^+\pi^-})} },
\end{eqnarray}
where we have used the calculation of the form
factors from Refs.~\cite{Lu:2007sg,Li:2008tk}.  The uncertainties shown in the
parenthesis  are from the nonfactorizable contributions and have
been conservatively taken as  large as $50\%$. The decay branching fraction
of $f_X\to \pi^+\pi^-$ is an important input in the analysis and we
have used $10\%$ for illustration.

{\it Comparison with theory.} For illustration, we will consider two widely-discussed
mixing  mechanisms of the scalar mesons and for an overview of alternative schemes   see  Refs.~\cite{note_on_scalar_mesons_below_2GeV,Amsler:2004ps,Klempt:2007cp} and many references therein. Because the decay width of
the $f_0(1500)$ is not compatible with the ordinary $\bar qq$ state,  it
is claimed that $f_0(1500)$ is primarily a scalar
glueball~\cite{Close:2000yk}, and  the mixing matrix through fitting
the data of two-body decays of scalar mesons is
\begin{eqnarray}
 \left( \begin{array}{c}
 f_0(1710) \\f_0(1500)\\ f_0(1370)
 \end{array} \right)
 = \left( \begin{array}{ccc} 0.36 & 0.93&0.09 \\
 -0.84 &0.35 &-0.41  \\
 0.40 &-0.07 &-0.91\end{array}\right)
\left(\begin{array}{c} G \\ \bar ss \\ \bar nn\end{array}
\right).\label{2}
\end{eqnarray}
Based on the SU(3) assumption for scalar mesons and the quenched
Lattice QCD results \cite{Chen:2005mg}, Cheng et
al.~\cite{Cheng:2006hu} reanalyzed all existing experimental data
and derived  the mixing coefficient matrix as
\begin{eqnarray}
 \left( \begin{array}{c}
 f_0(1710) \\f_0(1500)\\ f_0(1370)
 \end{array} \right)
 = \left( \begin{array}{ccc} 0.93 & 0.17&0.32 \\
 -0.03 &0.84 &-0.54  \\
 -0.36 &0.52 &0.78\end{array}\right)
\left(\begin{array}{c} G \\ \bar ss \\ \bar nn\end{array}
\right).\label{3}
\end{eqnarray}
Here, the $f_0(1710)$ tends to be a glueball. This is very
different from the first matrix of mixing coefficients in
Eq.(\ref{2}), while both  schemes can well explain the  data on the
production in $J/\psi$ and decays of the $f_0$.

The mixing scheme of Eq.~(\ref{3}) predicts a much larger production
branching ratio for $J/\psi\to \gamma f_0(1710)$ than $J/\psi\to
\gamma f_0(1500)$ and implies a relatively pure glueball around 1.7~GeV. 
In contrast, the mixing scheme of Eq.~(\ref{2}) suggests that
those two nearby states $f_0(1500)$ and $f_0(1710)$ both have
sizeable glueball components. It can be understood that the
interferences between the glueball and $\bar{q}q$ production would
lead to an enhanced production rate for $J/\psi\to \gamma
f_0(1710)$, but a suppression of $J/\psi\to \gamma f_0(1500)$. Such
an ambiguity reflects the lack of knowledge on the
glueball-$\bar{q}q$ coupling in the scalar sector. Qualitatively
speaking, it is strongly model-dependent to determine the glueball
component of scalar mesons in their strong productions and strong
decays. In this sense, it is interesting to recognize the advantages
of probing the flavor components of the scalar mesons in $B$ decays.
To be more specific, in  the decay of $B_s\to J/\psi f_0$ the mixing
coefficients in the second column, $\alpha_2$'s defined in Eq.~(\ref{eq:mixing_coefficient}),  
will be projected
out by the weak transitions. Thus, they provide a natural filter of
the $\bar ss$ component.

Apart from the measurements on $B_s\to J/\psi \pi^+\pi^-$,  both the
LHCb and D0 collaborations  have measured the branching ratio  of
the process  $B_s\to J/\psi
K^+K^-$~\cite{Aaij:2011ac,Abazov:2012dz}, in which no significant
evidence for any scalar resonance decaying into $K^+K^-$ is found.
Thus, it may be hard to interpret the $f_0(1710)$ as an  $\bar ss$
state since in this case the  $f_0(1710)$ mainly decays into $K^+K^-$.
From this viewpoint, it seems that the mixing in Eq.(\ref{2}) is
less favored compared to the one in  Eq.(\ref{3}), where the
production of the glueball dominated $f_0(1710)$ is expected to be
suppressed.

The present experimental precision does not allow a conclusion for
the $f_0(1500)$. Although in the mixing scheme of Eq.(\ref{3}), the
$f_0(1500)$ is favored to be produced via its $\bar ss$ component,
its decay branching ratio to $K^+K^-$ is relatively small, i.e.
$(8.6\pm 1.0)\%$~\cite{Beringer:1900zz}. With higher statistics
available in the future, a determination of the relative production
rates for the $f_0(1710)$ and the $f_0(1500)$ in $B_s\to J/\psi \pi^+\pi^-$
should be able to provide  crucial information about their
internal structure.

Supposing that the absence of $f_0(1710)$ in $K^+K^-$ is indeed due
to the dominance of an internal glueball component, one notices that
the such a scenario is consistent with the  recent Lattice QCD
calculation of Ref.~\cite{Gui:2012gx}, where the $f_0(1710)$ as a glueball
candidate would have a large production rate in $J/\psi$ radiative
decays, i.e. $J/\psi\to \gamma f_0(1710)$. It is also in agreement with the coupled channel study of the S-waves meson-meson scattering~\cite{Albaladejo:2008qa}, in which the $f_0(1710)$ and a
pole at 1.6 GeV, which is an important contribution to
the $f_0(1500)$, are identified as the scalar glueballs.

Regarding the $f_X$ resonance  discovered by Belle and LHCb,  we
explore two cases since the masses and widths of both $f_0(1370)$
and $f_0(1500)$ are close to the experimental values:

i) The  $f_X$ can be the $f_0(1500)$. From the PDG tables~\cite{Beringer:1900zz},
we quote
\begin{eqnarray}
 {\cal B}(f_0(1500)\to \pi^+\pi^-) &=& \frac{2}{3}\times  34.8\% = 23.2\%,
\end{eqnarray}
which implies
\begin{eqnarray}
 |\alpha_2| &=& (0.18\pm 0.09)   \times
\frac{F_1^{B\to K}(m_{J/\psi}^2)}{0.53}      \nonumber\\
&& \times 
\frac{1.22}{F_1^{B_s\to f_0(\bar ss)}(m_{J/\psi}^2)} \label{eq:alpha_2_Belle}.
\end{eqnarray}
Such a small value seems to favor the mixing matrix shown in
Eq.(\ref{2}).

ii)  The $f_X$ can be the $f_0(1370)$. The PDG quote that the $f_0(1370)\to
\rho\rho$ is its main decay mode. Both WA102~\cite{Barberis:2000cd}
and BES-II~\cite{Ablikim:2004wn} found that the branching ratio
fraction of $f_0(1370)\to \pi^+\pi^-$ over $f_0(1370)\to K^+K^-$ is
small, i.e. $\sim 20\%$~\cite{Ablikim:2004wn}. If this is the case,
the extracted coefficient $\alpha_2$ is of a similar size as the
value in Eq.~(\ref{eq:alpha_2_Belle}).  In such a situation, it
seems hard to distinguish the mixings given in Eq.~(\ref{2}) and
Eq.~(\ref{3}).

{\it Future improvements.} Although the present experimental status
does not allow us to make a definite conclusion on the $f_X$, we 
would expect that the situation will be greatly improved in the future. 
As we have shown above, the measurement of branching ratios of the 
$B_s\to J/\psi f_0$ with
high statistics will be able to pin down the flavor components of
the scalars. Therefore, a precise measurement of the relative
production rates of all (or some of) those scalars in $B_s\to J/\psi
f_0$ will be an ideal $\bar ss$ filter for the determination of the
$\bar ss$ components inside those scalar mesons. It is also possible
to use the $B\to J/\psi f_0$ decays as a flavor filter for the
non-strange $\bar qq$ components similar to that in $B_s\to J/\psi
f_0$. A combined measurements of $B_s\to J/\psi f_0$ and $B\to J/\psi
f_0$ will be very selective to scalar mixing models and can be
compared with the scalar production mechanisms studied in e.g.
$J/\psi\to \gamma f_0$.


Generically the branching fractions of the $b\to \bar ccd$ processes
are suppressed by  $V_{cd}^2/V_{cs}^2\sim 0.04$, which we suppose
would be compensated by the large luminosity of the future
experiments. The Super KEKB factory is expected to gather about 50 $ab^{-1}$
of data, which is two orders of magnitude larger than the data sample
collected on the KEKB collider~\cite{Aushev:2010bq}. With such a
high statistic data base, one might gain access to $B_s/B_d\to
J/\psi f_0( \gamma\gamma)$ in which  the scalar meson is
reconstructed in the two-photon final state. Compared to the
$B_s/B\to J/\psi f_0( \pi^+\pi^-, K^+K^-)$, in which  the decay of
the $f_0$ is not under control due to the unknown contributions from the
glueball, the $B_s/B_d\to J/\psi f_0( \gamma\gamma)$ is cleaner.
Due to the fact that the gluons are free of electromagnetic
interaction, the glueball component will not contribute. Thus, the
decay matrix elements  of the three $f_0$s  can be  determined by
the mixing coefficients and electric charges of the flavor
components.


{\it Conclusion.}  To summarize, using the available experimental
data we have demonstrated that the flavor SU(3) symmetry for the
light quarks holds quite well and can be applied to the study of $B$
decays into a scalar meson plus a charmonium. Our analysis suggests
that such a process would serve as an ideal flavor filter for
probing the quark contents of  scalar mesons. In the presence of
mixings between a glueball and ordinary $\bar qq$ mesons, we show that
the mixing parameters can be extracted and explicitly related to
experimental data from the LHCb, Belle and D0 collaborations.
Although the present experimental data sample does not allow a solid
conclusion on all those states, we have shown the sensitivities of
such a probe to the scalar structures. The future Super KEKB factory would
allow access to establishing the mixing pattern among those three
scalars, and possibly allow one to disentangle the long-standing puzzle
concerning the existence and mixings of the scalar
glueball predicted by QCD.

{\it Acknowledgements:} W.W. thanks Y.L. Shen for useful discussions and Institute of High Energy Physics and Tianjin University for their hospitalities  during his visit when part of this work was done. 
This work is supported in part by the DFG and the NSFC through funds
provided to the Sino-German CRC 110 ``Symmetries and the Emergence
of Structure in QCD'', the ``EU I3HP Study of Strongly Interacting
Matter''  under the Seventh Framework Program of the EU,  and the
NSFC under the Grant Nos. 11228512, 11235005, 11075168, and
11035006.






\begin{thebibliography}{11}


\bibitem{Bali:1993fb}
  G.~S.~Bali, {\it et al.} 
                  [UKQCD Collaboration],
  Phys.\ Lett.\  B {\bf 309}, 378 (1993);


  H.~Chen, J.~Sexton, A.~Vaccarino and D.~Weingarten,
  Nucl.\ Phys.\ Proc.\ Suppl.\  {\bf 34}, 357 (1994);


  C.~J.~Morningstar and M.~J.~Peardon,
  Phys.\ Rev.\  D {\bf 60}, 034509 (1999)
  ;


  A.~Vaccarino and D.~Weingarten,
  Phys.\ Rev.\  D {\bf 60}, 114501 (1999)
  ;

  C.~Liu,
  Chin.\ Phys.\ Lett.\  {\bf 18}, 187 (2001)
  ;

  D.~Q.~Liu, J.~M.~Wu and Y.~Chen,
  High Energy Phys.\ Nucl.\ Phys.\  {\bf 26}, 222 (2002)
  ;


  N.~Ishii, H.~Suganuma and H.~Matsufuru,
  Phys.\ Rev.\  D {\bf 66}, 014507 (2002);
 Phys. Rev. D {\bf 66}, 094506 (2002);


  M.~Loan, X.~Q.~Luo and Z.~H.~Luo,
  Int.\ J.\ Mod.\ Phys.\  A {\bf 21}, 2905 (2006). 

\bibitem{Chen:2005mg}
  Y.~Chen {\it et al.},
  Phys.\ Rev.\  D {\bf 73}, 014516 (2006).


\bibitem{He:2006qk} 
  X.~-G.~He and T.~-C.~Yuan,
  hep-ph/0612108; 
  C.~-H.~Chen and T.~-C.~Yuan,
  Phys.\ Lett.\ B {\bf 650}, 379 (2007).


\bibitem{Li:2011pg}
  J.~Li {\it et al.}  [Belle Collaboration],
  Phys.\ Rev.\ Lett.\  {\bf 106}, 121802 (2011)
  [arXiv:1102.2759 [hep-ex]].

\bibitem{LHCb:2012ae}
  RAaij {\it et al.}  [LHCb Collaboration],
  Phys.\ Rev.\ D {\bf 86}, 052006 (2012)
  [arXiv:1204.5643 [hep-ex]].



\bibitem{Aaij:2011ac}
  R.~Aaij {\it et al.}  [LHCb Collaboration],
  Phys.\ Rev.\ Lett.\  {\bf 108}, 151801 (2012)
  [arXiv:1112.4695 [hep-ex]].

\bibitem{Abazov:2012dz}
  V.~M.~Abazov {\it et al.}  [D0 Collaboration],
  [arXiv:1204.5723 [hep-ex]].


\bibitem{Close:2000yk}
  F.~E.~Close and A.~Kirk,
  Phys.\ Lett.\  B {\bf 483}, 345 (2000)
  [arXiv:hep-ph/0004241];

  F.~E.~Close and Q.~Zhao,
  Phys.\ Rev.\  D {\bf 71}, 094022 (2005)
  [arXiv:hep-ph/0504043].


\bibitem{Cheng:2006hu}
  H.~Y.~Cheng, C.~K.~Chua and K.~F.~Liu,
  Phys.\ Rev.\  D {\bf 74}, 094005 (2006)
  [arXiv:hep-ph/0607206].


\bibitem{note_on_scalar_mesons_below_2GeV}
C. Amsler, S.
Eidelman, T.
Gutsche, C. Hanhart, S. Spanier, and N.A.
T{\"o}rnqvist, NOTE ON SCALAR MESONS BELOW 2 GEV, review published in Particle Data Group~\cite{Beringer:1900zz}. 


\bibitem{Colangelo:2010wg}
  P.~Colangelo, F.~De Fazio and W.~Wang,
  Phys.\ Rev.\ D {\bf 83}, 094027 (2011);  Phys.\ Rev.\ D {\bf 81}, 074001 (2010). 



\bibitem{Beringer:1900zz}
  J.~Beringer {\it et al.}  [Particle Data Group Collaboration],
  Phys.\ Rev.\ D {\bf 86}, 010001 (2012).


\bibitem{LHCb:2012cw}
  RAaij {\it et al.}  [LHCb Collaboration],
  arXiv:1210.2631 [hep-ex].


\bibitem{Chang:2012gnb}
  M.~C.~Chang {\it et al.}[Belle Collaboration],
  Phys.\ Rev.\ D {\bf 85}, 091102 (2012)
  [arXiv:1203.3399 [hep-ex]].

\bibitem{Wang:2009rc}
  W.~Wang, Y.~-L.~Shen and C.~-D.~Lu,
  J.\ Phys.\ G {\bf 37}, 085006 (2010)
  [arXiv:0908.2216 [hep-ph]].


%


\bibitem{Li:2008tk}
  R.~-H.~Li, C.~-D.~Lu, W.~Wang and X.~-X.~Wang,
  Phys.\ Rev.\ D {\bf 79}, 014013 (2009)
  [arXiv:0811.2648 [hep-ph]].

\bibitem{Lu:2007sg}
  C.~D.~Lu, W.~Wang and Z.~T.~Wei,
  Phys.\ Rev.\  D {\bf 76}, 014013 (2007)
  [arXiv:hep-ph/0701265].







\bibitem{Amsler:2004ps} 
  C.~Amsler and N.~A.~Tornqvist,
  Phys.\ Rept.\  {\bf 389}, 61 (2004).

\bibitem{Klempt:2007cp} 
  E.~Klempt and A.~Zaitsev,
  Phys.\ Rept.\  {\bf 454}, 1 (2007)
  [arXiv:0708.4016 [hep-ph]].


\bibitem{Gui:2012gx}
  L.~-C.~Gui, Y.~Chen, G.~Li, C.~Liu, Y.~-B.~Liu, J.~-P.~Ma, Y.~-B.~Yang and J.~-B.~Zhang,
  arXiv:1206.0125 [hep-lat], to appear in Phys.Rev.Lett.


\bibitem{Albaladejo:2008qa} 
  M.~Albaladejo and J.~A.~Oller,
  Phys.\ Rev.\ Lett.\  {\bf 101}, 252002 (2008)
  [arXiv:0801.4929 [hep-ph]].
  
\bibitem{Barberis:2000cd}
  D.~Barberis {\it et al.}  [WA102 Collaboration],
  Phys.\ Lett.\ B {\bf 479}, 59 (2000)  [hep-ex/0003033].

\bibitem{Ablikim:2004wn}
  M.~Ablikim {\it et al.}  [BES Collaboration],
  Phys.\ Lett.\ B {\bf 607}, 243 (2005)  [hep-ex/0411001].




\bibitem{Aushev:2010bq}
  T.~Aushev, W.~Bartel, A.~Bondar, J.~Brodzicka, T.~E.~Browder, P.~Chang, Y.~Chao and K.~F.~Chen {\it et al.},
  arXiv:1002.5012 [hep-ex].
\end{thebibliography}
\end{document}